\documentclass[twocolumn,floats]{revtex4}

\usepackage{graphics,graphicx,epsfig}
\usepackage{bm}

\begin{document}
\title{A New Model for Collective Behaviors of Animals}
\author{P.~The~Nguyen$^a$, V.~Thanh~Ngo$^b$\footnote{Corresponding author, E-mail: nvthanh@iop.vast.ac.vn } and H.~T.~Diep$^c$}
\address{$^a$Department of  Natural Science, Duytan University,\\  K7/25 Quang Trung, Haichau, Danang, Vietnam\\
$^b$Institute of Physics - Vietnam Academy of Science and Technology,\\ 10 Daotan, Ngoc Khanh, Badinh, Hanoi, Vietnam\\
$^c$Laboratoire de Physique Th\'{e}orique et Mod\'{e}lisation, Universit\'{e} de Cergy-Pontoise, CNRS,\\
UMR 8089, 2 Avenue Adolphe Chauvin, F-95302 Cergy-Pontoise Cedex, France}

\begin{abstract}
We propose a new model in order to study behaviors of self-organized system such as a group of animals. We assume that the individuals have two degrees of freedom corresponding one to their internal state and the other to their external state. The external state is characterized by its moving orientation. The rule of the interaction between the individuals is determined by the internal state which can be either in the non-excited state or in the excited state. The system is put under a source of external perturbation called ``noise". To study the behavior of the model with varying noise, we use the Monte-Carlo simulation technique. The result clearly shows two first-order transitions separating the system into three phases: with increasing noise, the system undergoes a phase transition from a dilute disordered phase to an ordered compact phase and then to the disordered dispersed phase. These phases correspond to behaviors of animals: uncollected state at low noise, flocking at medium noise and runaway at high noise, respectively.
\end{abstract}

\maketitle

\section{INTRODUCTION}

The collective behavior of animals is a widely observed phenomenon in various biological systems. It is one of the main topics which have been extensively investigated during the last two decades using methods borrowed from many different areas of science including physics, applied mathematics and engineering.  For recent papers, the reader is referred to works of Couzin and coworkers, of Albano and his group among others \cite{Couzin2009,Albano,Aldana,Saracco,Chate}.  In these works, many aspects of animal groups have been discussed, among which the collective motion and their origin, the relation between individual and collective behaviors and the relation between group size and collective decision-making.

The flocking is a behavior of some animal species where they stay together in a group for social reasons. They derive many benefits from this behavior including defence against predators, easier collective moving, enhanced foraging success and higher success in finding a mate. When they are faced with a danger such as predators, their natural instinct is to flee not to fight. They use their natural herding instinct to bind together in a group for safety. All individuals of the group will move away from the predator in the same direction and then stampede as fast as they can when being under the predator's attack. If there is no danger, then they are spread to find foods instead of staying in the flocking state.
Many experimental facts and observations have been for example mentioned in the review of Vicsek and Zafeiris \cite{Vicsek2012}. Well-known examples are found in populations such as large schools of fish~\cite{Huth} or gatherings of birds~\cite{Maldonado}. Biologically, it is known that the flocking behavior is advantageous for survival of a population~\cite{Werner,Pitcher,Cresswell}: reducing the risk of capture by predators, increasing higher mating efficiency, easier search for food, efficient learning of external stimuli, and reducing overall aggression~\cite{Cashing,Parrish,Adioui,Zheng}.

In 1987, Reynolds first suggested a simple model consisting of three rules: separation, alignment, and cohesion rules~\cite{Reynolds}. These rules describe the behavior of each individual in interaction with other neighboring individuals. All or some of the three rules were mathematically expressed and then analyzed by Vicsek and his coworkers~\cite{vicsek95,Csahok,vicsek969900,Vicsek2012}. They mainly focused on the transition between coherently moving and runaway in a stampede.

The flocking behavior has been conventionally studied through simulation in two frameworks: population (Eulerian or continuum models) and individuals (agents or particle-based models)~\cite{Adioui,Grunbaum,Inada}. In the population framework, the flock was collectively addressed while flock-density was used as a key variable to present spatial and temporal dynamics of aggregation frequently with partial differential equations of advection-diffusion reaction~\cite{Mogilner,Murray}. In the individual framework, the flock of agents has been simulated by using ordinary and stochastic equations of motion to describe interactions among agents~\cite{vicsek95,Huth,Niwa,Tu,Toner}. This approach attempted to replicate naturally observed phenomena from not only animal groups but also other self-propelled characteristics~\cite{Spector} and to compare the evolved characteristics with those of actual animal flocking~\cite{Wood} in order to better understand  possible mechanisms by which these characteristics may have evolved.

The previous models have predicted with success flocking behavior of animal groups at high noise. They have, however, difficulties to explain uncollected states at very low noise.  This has motivated our present work: in addition to the interaction between neighboring animals, we introduce an internal degree of freedom which indicates whether an animal is in a non-excited or an excited state, we will show that the collective behavior of animals in the whole range of noise can be explained.

Section II is devoted to the description of the model. Section III shows the Monte Carlo (MC) simulation results for testing our model. Conclusions are given in Sec. IV.

\section{The Model}

In biology, all the members of an animal group are spread to find foods if there is no danger. In this situation, they are distributed in the space with a small concentration and out of alignment. So, we say the group of animals is in a ``uncollected" behavior. When the animals are faced with danger such as predators, they bind together in a small area for safety with the same orientation and high concentration. This state is called ``flocking" state. Facing a danger, animals will move away from the predator in the same direction and then stampede as fast as they can when being under the predator's attack. At the final stage, they are in a ``runaway" state.

In order to study the phase transition behavior, we have to map the group of animals into a physical system. We consider an animal as a particle $i$ with two degrees of freedom:  an external parameter $\sigma_i$ characterizing the animal orientation which depends on its interaction with the others, and an internal parameter $S_i$ indicating either it is in the non-excited ($S_i = 0$) or in the excited ($S_i = 1$) individual state.

Let us define the internal degree of freedom. The internal state of a particle can be described by two levels: the first level is characterized by a negative energy $-\epsilon$ which describes the calm, unworried stable state, and the second level by a positive energy $+\epsilon$ which expresses some degree of anxiety.  Introducing an external noise $\eta$ which plays the role of the temperature in statistical physics, we can write the canonical probability of these two states as \cite{diepbook15}
\begin{equation}
p_1=\frac{\mbox{e}^{\epsilon/\eta}}{z} \ \ , \  p_2=\frac{\mbox{e}^{-\epsilon/\eta}}{z}
\end{equation}
where $z$ is the partition function given by
\begin{equation}
z=\mbox{e}^{\epsilon/\eta}+\mbox{e}^{-\epsilon/\eta}=2\cosh(\epsilon/\eta)
\end{equation}
Note that $p_1>p_2$ because $\epsilon>0$. So,  the ``positive" tendency towards tranquility
is proportional to $p_1-p_2$. The total number of particles having this tranquility state which is called non-excited state is thus given by
\begin{eqnarray}
N_0&=&N(p_1-p_2)=N \left[\frac{\mbox{e}^{\epsilon/\eta}-\mbox{e}^{-\epsilon/\eta}}{z}\right]\nonumber\\
&=&N\tanh(\epsilon/\eta)\label{eq:nexcited}
\end{eqnarray}
where $N$ is the total number of particles in the system.
The number of excited particles is then
\begin{equation}\label{NE}
N_e=N-N_0
\end{equation}
One sees in the above equations that in zero noise one has $N_0=N$, namely there is as expected no excited particles. While, when the noise goes to infinity, one has $N_0=0$ and $N_e=N$: all particles are excited. At a given noise, statistically one has a well-defined $N_0$ and $N_e$ with the conservation $N=N_0+N_e$.

In the simulation which will be done below at a given noise, we generate at random $N_0$ according to Eq. (\ref{eq:nexcited}). The remaining particles in the system are thus in the excited state.

Now, we put this system of individuals on the lattice where each individual can move on 2D triangular lattice of linear size $L$. The number of lattice sites should be greater than that of individuals, i.e., $L^2 \gg N$. We denote by $\sigma_i$ the orientation of individual $S_i$: $\sigma_i$ is defined as in a $q$-state Potts model, i.e. $\sigma_i = 1,2,\ldots, q$. For simplicity, we consider $q=6$, so the orientations $\sigma_i= 1,2,\ldots, 6$ of individual $S_i$ can be defined by the vectors which connect a site to its nearest neighbors (NN) with the following angles measured from the $x$ axis: $\varphi_i = 0,\pi/3,\ldots, 5\pi/3$. The interaction between nearest-neighboring animals is given by the Hamiltonian
\begin{equation}
\label{eq:hamil}
{\mathcal H}= \sum_{<i,j>} K_{i,j} \cos[\pi(\sigma_i-\sigma_j)/3],
\end{equation}
where the sum $\sum_{<i,j>}$ is made over up to the third nearest neighboring individuals $S_i$ and $S_j$. For simplicity, $K_{i,j}$ is assumed to take the form of the Lennard-Jones potential:
$$
K_{i,j} = 4 J\left[(r_0/r_{i,j})^{12}-(r_0/r_{i,j})^6\right],
$$
where $r_{i,j}$ is the distance between two individuals. We choose $r_0=0.89$ in order that $K_{i,j}\simeq J$ at $r_{i,j} = 1$ which is the lattice spacing, $J$ being a positive constant. The interaction between two individuals depends on their internal state: $K_{i,j}= 0$ if both individuals are non-excited $S_i = S_j =0$, and $K_{i,j}\neq 0$ if otherwise.
We emphasize that the interaction rule we impose means that an excited particle can interact with non-excited particles, dragging them into a collective motion.

Let us note that, without the internal degree of freedom and without particle motions,  the above Hamiltonian is the localized Potts clock model  which has been solved for $q=2$, 3 and 4 \cite{Baxter}. The case of very large $q$ has been solved by Fr\"{o}hlich and Spencer \cite{Frohlich}. However, for $q=5$, 6, ... there are not (yet) exact results. Of course, our model is more complicated because the particles are mobile with an internal degree of freedom.

\section{The phase-transition behavior}
 We use the MC simulation technique to study the above model. The main physical quantities such as the order parameter $Q$ and the concentration $\rho$ are defined in what follows.

For $Q$, we have

\begin{equation} \label{eq:aq}
Q =  \frac{qM_{\mbox{max}}-1}{q-1}
\end{equation}
where
\begin{equation}
M_{\mbox{max}}=\frac{\mbox{max}(M_1,M_2,...,M_q)}{N}
\end{equation}
with
\begin{equation}
M_{i}=\sum_j \delta_{\sigma_j,i}  \ \ (i=1,2,...,q)
\end{equation}
where the sum on $j$ is performed over all sites of the system. One
sees that in the ordered state where there is only one kind of
$\sigma_j$ one has $Q=1$. In the disordered phase where all values
of $\sigma_j$ are equally present, namely $M_{i}=N/q$ for any $i=1,...,q$,
one has $Q=0$.

For the concentration $\rho$ we have

\begin{equation}
       \rho = \frac{1}{N}\sum_{i=1}^N n_{i}       \label{eq:ar},
\end{equation}
where $n_i$ is the number of NN individuals around $S_i$.
The quantity $\rho$ in Eq.~(\ref{eq:ar}) characterizes the spatial distribution of the individuals. The behavior of animals can be adequately described by the two parameters $Q$ and $\rho$.

 In the simulations, we use $\epsilon =0.04$, $J = 1.0$ (taken as the unit of energy) and $N = 100$, $400$ and $900$ with the lattice size $L = 4\times \sqrt{N}$. For the initial configuration of the system, we randomly generate the particle positions on the lattice and take orientations of all individuals from an uniform distribution. At each MC step,  we randomly choose $N_0$ individuals (according to Eq.~(\ref{eq:nexcited})), and set them to be non-excited $S_i = 0$. The remaining particles are in the excited state $S_i = 1$. Their position and orientation are updated by Metropolis algorithm.
 At each $\eta$, the equilibration time lies around $4\times 10^6$ MC steps per individual and we compute statistical averages over $8\times 10^6$ MC steps per individual. Each time a particle gets out of the lattice at one boundary, we put it into the system by the other end using the periodic boundary conditions in order to conserve the total number of particles.

\begin{figure}[!t]
\includegraphics[width=2.5in]{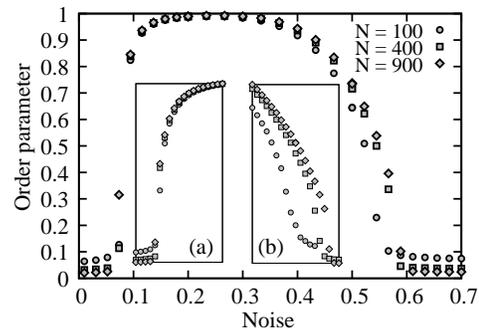}%
\caption{Order parameter versus noise with the system sizes $N=100$ (circles), 400 (squares) and 900 (diamonds). The insets show the enlarged scale at low (a) and high (b) noise.
\label{fig:f1}}
\end{figure}

\begin{figure}[!t]
\includegraphics[width=2.5in]{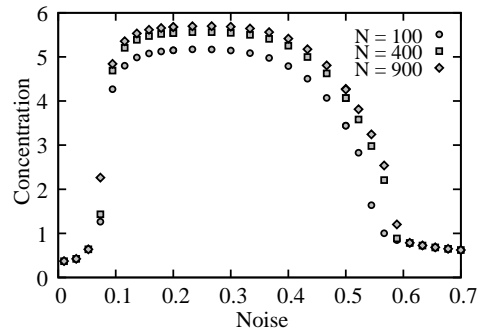}%
\caption{Concentration versus noise with $N=100$ (circles), 400 (squares) and 900 (diamonds).\label{fig:f2}}
\end{figure}

We plot in figures~\ref{fig:f1} and \ref{fig:f2} the order parameter and the concentration as a function of external noise for several system sizes. It clearly shows the existence of three phases which are separated by the two transitions at very low noise and high noise. In phase I at low noise, the system is in the disordered phase with low concentration ($Q\approx 0$ and $\rho\approx 0$). This phase is equivalent to the uncollected behavior of an animal group which is dispersed over the whole space because almost of them are non-excited, it is called ``free'' phase with a few ``contacts" between the individuals. Phase III, at high noise, is spatially sparse as the phase I, but the particles are very mobile in all directions. It describes the runaway behavior. At medium noise, the animals are in phase II where they are compactly moving in an ordered phase with $Q\approx 1$ and the maximum concentration $\rho\approx 6$. This phase corresponds to the flocking behavior.

\begin{figure}[!t]
\includegraphics[width=3.in]{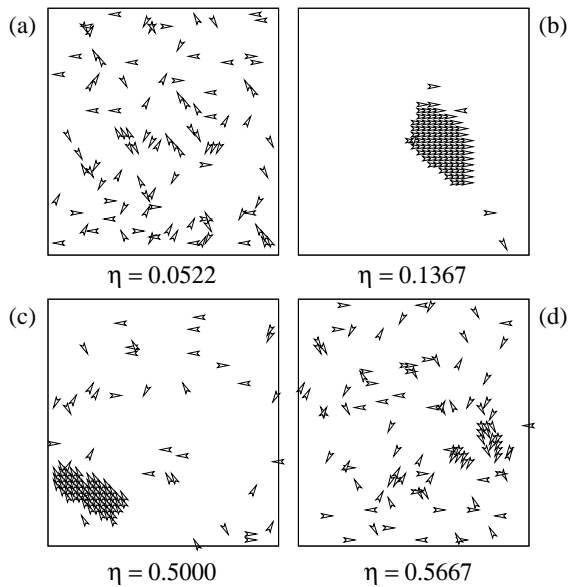}%
\caption{Snapshots at different values of noise with $N=100$. The arrow indicates the orientation of individual. See text for comments.}
\label{fig:f3}
\end{figure}

The spatial distribution of the individuals is also presented in Fig.~\ref{fig:f3} with snapshots taken at several values of noise $\eta = 0.0522, 0.1367, 0.5000$ and $0.5667$, at the final stage of a MC simulation. The vectors indicate their position and moving orientation. One sees that Figs~\ref{fig:f3}(a,d) show that the system has the same distribution where they are randomly distributed in the plane with different orientations. However, the instantaneous snapshots do not permit to see the difference between two phases: the particles in phase I are almost immobile while they are moving very fast in a disordered manner in phase III (this difference is seen in videos). Fig.~\ref{fig:f3}(b) shows the flocking behavior, and Fig.~\ref{fig:f3}(c) shows the distribution of the system at the noise closes to the II-III transition point.


\begin{figure}[!t]
\includegraphics[width=2.5in]{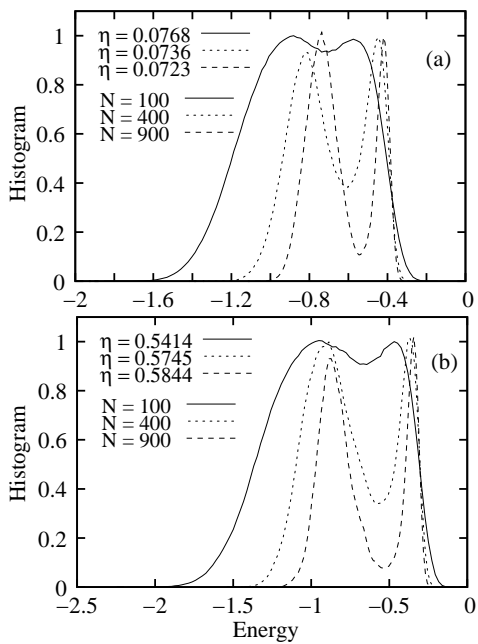}%
\caption{Histogram versus energy at low (a) and high (b) noise at the transition points, with the system sizes $N=100$ (solid), 400 (dotted) and 900 (dashed).}
\label{fig:f4}
\end{figure}

 At this stage, let us examine again Fig.~\ref{fig:f1}: it shows  discontinuities of the order parameter at both transitions I-II and II-III, indicating a signature of a first-order transition. This abrupt behavior corresponds to the fact that the animals immediately flock when they are faced with danger, and they runaway when being under a predator's attack. These behaviors are more likely a first-order transition. To confirm the first-order transition we have performed the energy histogram at the transition point. We show in Fig.~\ref{fig:f4} the energy histogram at two transitions I-II and II-III, for three system sizes. The double-peak histograms are clearly shown for both transitions at low noise (a) and high noise (b).
The dip between the two maxima becomes deeper with increasing size. Note that a ``true" discontinuity happens only when the dip comes down to zero. The distance between the two peaks is then the latent heat. To see this, we need sizes much larger than $N = 900$. But this is out of the scope of our present purpose.

\begin{figure}[!t]
\includegraphics[width=2.5in]{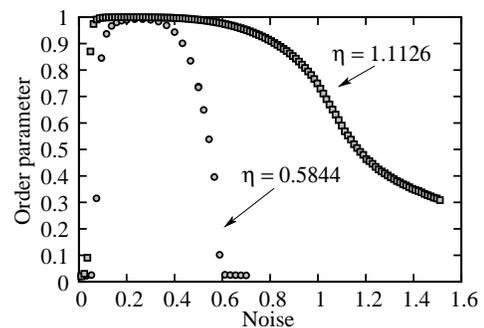}%
\caption{Order parameter versus noise with $N=100$ for the localized lattice model (diamonds) and lattice gas model (circles).}
\label{fig:f5}
\end{figure}

Let us discuss on the phase transition at high noise. It is known that the $q$-state  Potts model localized on a lattice has a first-order transition with $q \geq 4$  in two dimensions.  The clock Potts model localized on a lattice was solved only for $q\leq 4$ \cite{Baxter}, though it was solved for very large $q$ \cite{Frohlich}. The model treated in the present paper corresponds to a mobile Potts model where particles can go from one lattice site to another. This was never solved before.   Note however that the transition II-III corresponds not only to an orientational disordering of the Potts parameter but also to the breaking of the compactness of the flocking state. This is similar to a melting transition where the solid phase melts into the liquid phase. The melting has often a first-order character in three dimensions \cite{Gomez2003}. In two dimensions, it is known that long-range solid ordering does not survive at finite temperatures if the system has isotropic short-range interactions according to Nelson and Halperin \cite{NelsonHalperin}: at a first critical temperature, bound pairs of dislocations formed at low temperatures are unbound, giving rise to a phase with no translational ordering but with orientational hexatic structure. The latter phase undergoes a ``critical" phase transition to the disordered phase. It was  however found by MC simulation that the second transition is not critical but it is a first-order two-dimensional disclination melting \cite{Janke}.
 Our results shown in Figs. \ref{fig:f3} (c,d) and \ref{fig:f4} indicate that our model undergoes a first-order transition from the hexatic ordered phase to the liquid phase, in agreement with earlier MC simulations \cite{Janke}.

 Finally, to compare our mobile Potts model with the corresponding localized Potts model, we have simulated the same system defined by (\ref{eq:hamil}) including the internal degree of freedom on a lattice of size $L^2 = N$, i.e. the individuals are not allowed to move on the lattice. We show in Fig.~\ref{fig:f5} results of both mobile and localized Potts models: the transition temperature is about $0.5844$ for the mobile case, while it is $\eta \approx 1.1126$ for the localized Potts model.

\section{Conclusions}
We proposed in this paper a simple model for studying the behavior of animal groups as a function of an external perturbation, called ``noise", such as dangers coming from predators.
 We showed that with increasing noise, the system has three phases I, II and III separated by two transitions, the first transition occurs at a low noise and and the second one at high noise. The three phases are disordered, ordered and melted phases which correspond respectively to the following behaviors of animals: uncollected state, flocking state and runaway state. Both transitions from one phase to another are found to be of the first-order.

This model is similar to the lattice gas model developed by Csah\'{o}k {\it et. al.}~\cite{Csahok} but in our model the alignment rules of the individuals depend on the internal state, excited or non-excited, of its neighbors including itself. Therefore, our model can be applied not only for studying the phase transition behavior of animals at high noise as in previous models, but also for analyzing the animal behavior at low noise where animals are in a uncollected, disordered phase.

This work was supported by the Nafosted (Vietnam National Foundation for Science and Technology Development), Grant No. 103.02-2011.55.


\begin{thebibliography}{99}
\bibitem{Vicsek2012} T.~Vicsek and A.~Zafeiris, Phys. Reports \textbf{517}, 71 (2012).
\bibitem{Couzin2009} I.~D.~Couzin, Trends in Cognitive Sciences \textbf{13}, 36 (2009); I.~D.~Couzin, J.~Krause, N.~R.~Franks and S.~A.~Levin, Nature \textbf{433}, 513 (2005).
\bibitem{Albano} G.~Baglietto and E.~V.~Albano, Computer Physics Communications \textbf{180}, 527 (2009).
\bibitem{Aldana} M.~Aldana, H.~Larralde and B.~V\'asquez, Internat. J. of Modern Physics B \textbf{23}, 3459 (2009).
\bibitem{Saracco} G.~P.~Saracco, G.~Gonnella, D.~Marenduzzo and E.~Orlandini, Cent. Eur. J. Phys.  \textbf{10}, 1109 (2012).
\bibitem{Chate} H.~Chat\'{e}, F.~Ginelli, G.~Gr\'{e}goire, F.~Peruani and F.~Raynaud, Eur. Phys. J. B \textbf{64}, 451 (2008).
\bibitem{Huth} A.~Huth and C.~Wissel, Ecol. Modell., \textbf{75}, 135 (1994).
\bibitem {Maldonado}M.~Maldonado-Coelho and M.~A.~Marini, Biol. Conserv. \textbf{116}, 19 (2004).
\bibitem{Werner} G.~M.~Werner and M.~G.~Dyer, \textit{Evolution of herding behavior in artificial animals}, Proceedings of the Second International Conference on Simulation of Adaptive Behavior, MIT Press, Cambridge, MA (1992).
\bibitem{Pitcher} T.~J.~Pitcher (Ed.) and J.~K.~Parrish, Pitcher, \textit{The Behavior of Teleost Fishes}, (Chapman \& Hall, New York, 1993).
\bibitem{Cresswell} W.~Cresswell, Tringa tetanus. Anim. Behav. \textbf{47}, 433 (1994).
\bibitem{Cashing} D.~H.~Cashing and F.~R.~Harden-Jones, Nature \textbf{218}, 918 (1968).
\bibitem{Parrish} J.~K.~Parrish, S.~V.~Viscido, and D.~Gr\"{u}nbaum, Biol. Bull. \textbf{202}, 296 (2002).
\bibitem{Adioui} M.~Adioui, J.~P.~Treuil, and O.~Arino, Ecol. Modell. \textbf{167}, 19 (2003).
\bibitem{Zheng}  M.~Zheng, Y.~Kashimori, O.~Hoshino, K.~Fujita, and T.~Kambara, J. Theor. Biol. \textbf{235}, 153 (2005).
\bibitem{Reynolds} C.~W.~Reynolds, \textit{Flocks, herds, and schools: a distributed behavioral model}, Computer Graphics, SIGGRAPH'87 Conference Proceedings \textbf{21}, 25 (1987).
\bibitem{vicsek95} T.~Vicsek, A.~Czirok, E.~Ben-Jacob, I.~Cohen, O.~Shochet, Phys. Rev. Lett. \textbf{75}, 1226 (1995).
\bibitem{Csahok} Z.~Csah\'{o}k and T.~Vicsek, Phys. Rev. E \textbf{52}, 5297 (1995).
\bibitem{vicsek969900}A.~Czirok, H.~E.~Stanley, and T.~Vicsek, Phys. Phys. A, \textbf{30}, 137 (1996);
A.~Czir\'{o}k, M.~Vicsek, T.~Vicsek, Physica A \textbf{264}, 299 (1999);
A.~Czir\'{o}k and T.~Vicsek, Physica A \textbf{281}, 17 (2000).
\bibitem{Grunbaum} D.~Gr\"{u}nbaum and A.~Okubo, {\it Frontiers in Mathematical Biology}, (S.A. Levin (Ed.), Springer, Berlin, 1994).
\bibitem{Inada} Y.~Inada and K.~Kawachi, J. Theor. Biol. \textbf{214}, 371 (2002).
\bibitem{Mogilner} A.~Mogilner and L.~Edelstein-Keshet, Physica D \textbf{89}, 346 (1996).
\bibitem{Murray} J.~D.~Murray, \textit{Mathematical Biology}, (vols. 1 and 2, Springer, New York, 2003).
\bibitem{Niwa} H.~S.~Niwa, J. Theor. Biol. \textbf{181}, 47 (1996).
\bibitem{Tu} Y.~Tu, Phys. A \textbf{281}, 30 (2000).
\bibitem {Toner} J.~Toner and Y.~Tu, Phys. Rev. E \textbf{58}, 4828 (1998).
\bibitem{Spector} L.~Spector, J.~Klein, C.~Perry, and M.~Feinstein, GPEM \textbf{6}, 111 (2005).
\bibitem{Wood} A.~J.~Wood and G.~J.~Ackland, P. Roy. Soc. B-Biol. Sci., \textbf{274}, 1637 (2007).
\bibitem{diepbook15} H.~T.~Diep, \textit{Statistical Physics: Fundamentals and Application to Condensed Matter, Lectures, Problems and Solutions} (World Scientific, Singapore, 2015), chapter 3.
\bibitem{Baxter} R.~J.~Baxter, \textit{Exactly Solved Models in Statistical Physics}, (Academic Press, California, USA, 1982).
\bibitem{Frohlich}J.~Fr\"{o}hlich and T.~Spencer, Communications in Mathematical Physics \textbf{81}, 527-602 (1981).
\bibitem{Gomez2003} L.~G\'{o}mez, A.~Dobry, C.~Geuting, H.~T.~Diep and L.~Burakovsky, Phys. Rev. Lett. \textbf{90}, 095701 (2003) and references therein.
\bibitem{NelsonHalperin} D.~R.~Nelson and B.~I.~Halperin, Phys. Rev. B \textbf{19}, 2457 (1979).
\bibitem{Janke} W.~Janke and H.~Kleinert, Phys. Letters \textbf{105A}, 134 (1984).
\end{thebibliography}
\end{document}